\documentclass{appolb}
\usepackage{graphicx}

\begin{document}

\title{A comparison study of slow--subdiffusion and subdiffusion%
\thanks{Presented at 25th Marian Smoluchowski Symposium on Statistical Physics}%
}

\author{K.D. Lewandowska
\address{Department of Radiological Informatics and Statistics,\\ Medical University of
         Gda\'nsk,\\ ul. Tuwima 15, 80-210 Gda\'nsk, Poland.}
\and
Tadeusz Koszto{\l}owicz
\address{Institute of Physics, Jan Kochanowski University,\\
         ul. \'Swi\c{e}tokrzyska 15, 25-406 Kielce, Poland.}
}

\maketitle
\begin{abstract}
We study slow--subdiffusion in comparison to subdiffusion. Both of the processes are treated as random walks and can be described within  continuous time random walk formalism. However, the probability density of the waiting time of a random walker to take its next step $\omega(t)$ is assumed over a long time limit in the form $\omega(t)\sim 1/t^{\alpha+1}$ for subdiffusion, and in the form $\omega(t)\sim h(t)/t$ for slow--subdiffusion [$h(t)$ is a slowly varying function]. We show that Green functions for slow--subdiffusion and subdiffusion can be very similar when the subdiffusion coefficient $D_\alpha$ depends on time. This creates the possibility of describing slow--subdiffusion by means of subdiffusion with a small value of the subdiffusion parameter $\alpha$, and the subdiffusion coefficient $D_\alpha$ varying over time.
\end{abstract}
\PACS{05.40.-a, 05.60.-k, 05.40.Fb}

\section{Introduction}

Anomalous diffusion is usually described within a framework of the continuous time random walk in which the probability densiy of finding a random walker at time $t$ at point $x$, $P(x,t)$, depends on the probability density of the waiting time of the random walker to take its next step $\omega(t)$ and the probability density of jump length $\lambda(x)$. When $\omega(t)$ and $\lambda(x)$ are independent then the probability density $P(x,t)$ reads in terms of the Laplace transform $\mathcal{L}\left\{\omega(t)\right\}\equiv \hat{\omega}(s)\equiv\int_{0}^{\infty}e^{-st}f(t)dt$ and the Fourier transform $\mathcal{F}\left\{\lambda(x)\right\}\equiv\hat{\lambda}(k)\equiv\int_{-\infty}^{\infty}e^{ikx}f(x)dx$ \cite{mk}
\begin{equation}
  \label{eq_1}
  \hat{P}(k,s)=\frac{1-\hat{\omega}(s)}{s}\frac{1}{1-\hat{\omega}(s)\hat{\lambda}(k)}\;.
\end{equation}
As far as we know, an inverse transform of the above equation in the most general case has not been found yet, with the expection of a few very special cases. For this reason, Eq.~(\ref{eq_1}) is usually considered within the limit of small values of $s$ and $k$. For subdiffusion, it is assumed that  the first moment of $\omega(t)$ (the average value) is infinite and that the Laplace transfom $\hat{\omega}(s)$ has the following form for small values of $s$
\begin{equation}
  \label{eq_2}
  \hat{\omega}(s)\cong 1-\tau^{\alpha}s_{\alpha}\;,
\end{equation}
where subdiffusion parameter $\alpha$ obeys $0<\alpha<1$ and $\tau_\alpha$ is a positive parameter, whereas all the moments of the natural order of $\lambda(x)$ are finite. Assuming that $\lambda(t)$ is a symmetric function, the Fourier transform  $\hat{\lambda}(k)$ has the following form for small values of $k$
\begin{equation}
  \label{eq_3}
  \hat{\lambda}(k)\cong 1-\sigma^{2}\frac{k^2}{2}\;,
\end{equation}
where $\sigma^{2}$ is the second moment of $\lambda(x)$. Computing the inverse Laplace and Fourier transforms leads to $\omega(t)$ being proportional to $1/t^{1+\alpha}$ (therefore $\omega(t)$ is refered to as a heavy--tailed distribution) and $\lambda(x)$ in the form of Gauss distribution. A subdiffusion equation which is obtained as a result of continuous time random walk formalism is a linear differential equation with a fractional time derivative.

A random walk in which $\omega(t)$ is a superheavy--tailed distribution has been recently studied \cite{dk,dybkl}. The superheavy--tailed distribution means that 
\begin{equation}
  \label{eq_4}
  \omega(t)\sim\frac{h(t)}{t}\;,
\end{equation}
where $h(t)$ is a slowly varying function, i.e. $h(\chi t)/h(t)\rightarrow 1$ when $t\rightarrow +\infty$ for all positive values of $\chi$. All the fractional moments of $\omega(t)$ ($\langle \omega^c(t)\rangle=\int x^c\omega(t)dt$, $c>0$) are infinite. Such a process qualitatively differs from subdiffusion and is usually called slow--subdiffusion. It is rather difficult to find an equation with fractional derivatives which describes such a process since an order of the derivative would be very close to zero. 

In this paper we will compare Green functions for subdiffusion and slow--subdiffusion and we will also show that slow--subdiffusion could be described by subdiffusion with small values of subdiffusion parameter $\alpha$ and with subdiffusion coefficient $D_\alpha$ which depends on time.

\section{Green functions}

Green functions for subdiffusive transport in an unrestricted system is well known. They can be obtained after substituting Eqs.~(\ref{eq_2}) and (\ref{eq_3}) into Eq.~(\ref{eq_1}) and then computing the inverse Laplace and Fourier transforms. After calculation we obtain
\begin{equation}
  \label{eq_5}
  P(x,t)=\frac{1}{2\sqrt{D_\alpha}} f_{\alpha/2-1,\alpha/2}\left(t;\frac{|x|}{\sqrt{D_\alpha}}\right)\;,
\end{equation}
where \cite{koszt}
\begin{equation}
  \label{eq_6}
  f_{\nu,\beta}(t;a)=\frac{1}{t^{1+\nu}}\sum_{k=0}^{\infty} \frac{1}{k!\Gamma(-k\beta-\nu)}\left(-\frac{a}{t^\beta}\right)^k\;.
\end{equation}

The Green function for slow--subdiffusive transport in an unrestriced system was derived in \cite{dk} and reads over a long time limit
\begin{equation}
  \label{eq_7}
  P_{s}(x,t)=\sqrt{\frac{V(t)}{2\sigma^2}}e^{-|x|\sqrt{2V(t)/\sigma^2}}\;,
\end{equation}
where $V(t)$ is the complementary cumulative distribution function of waiting times $V(t)=1-\int_0^t\omega(t)dt=\int_t^\infty\omega(t)dt$, $V(0)=1$ and $V(+\infty)=0$. In the further consideration we take $\omega(t)$ in the following form
\begin{equation}
  \label{eq_7a}
  \omega(t)=\frac{(r-1)\ln^{r-1}\eta}{(\eta+t)\ln^r(\eta+t)}\;,
\end{equation}
where $r>1$ and $\eta>1$, which provides \cite{dk}
\begin{equation}
  \label{eq_8}
  V(t)=\left[\frac{\ln\eta}{\ln(\eta+t)}\right]^{r-1}\;.
\end{equation}

\section{Comparision between the Green functions for slow--subdiffusion and subdiffusion}

\begin{figure}[htb]
\centerline{%
\includegraphics[scale=0.4]{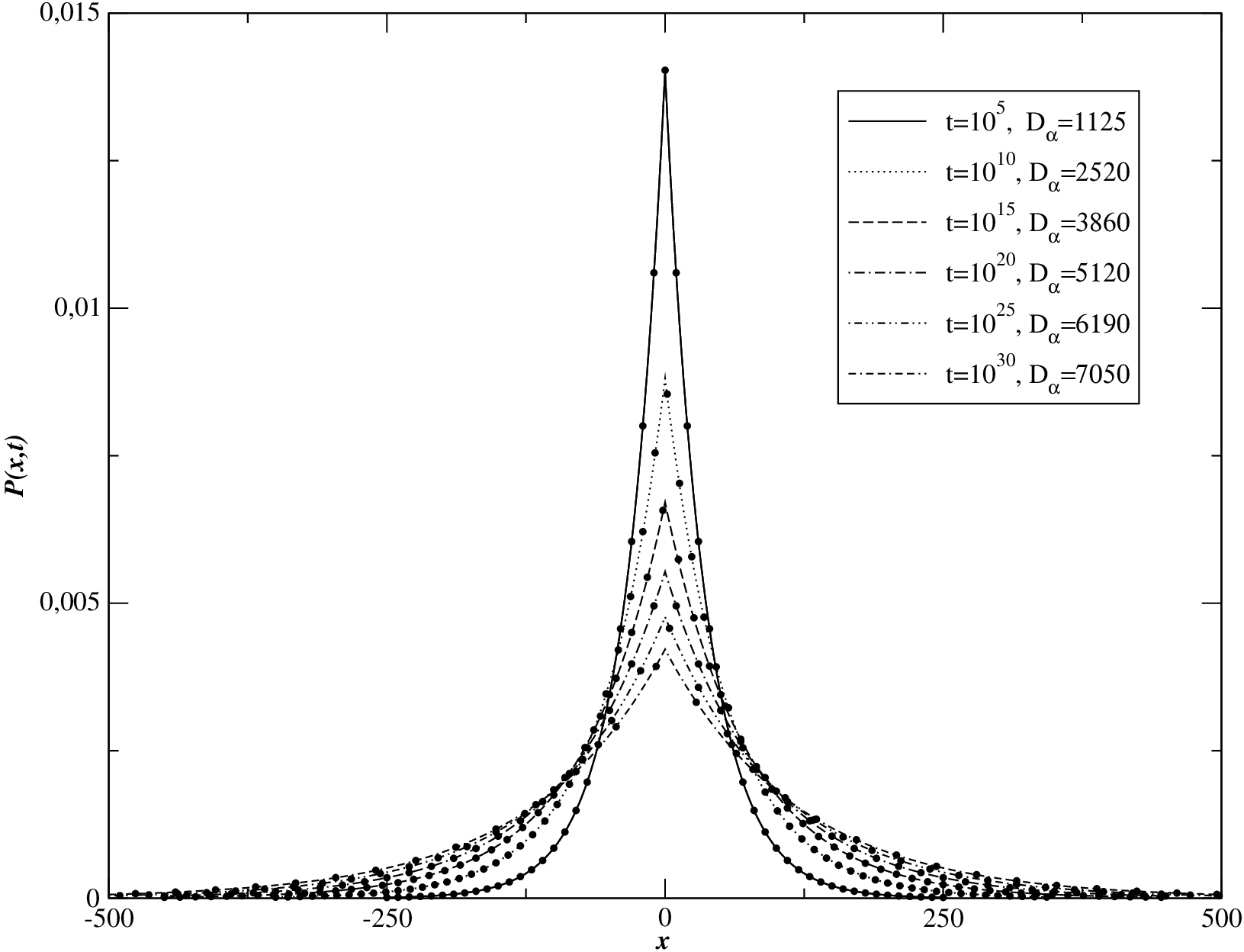}}
\caption{The comparision between Green functions for subdiffusion [Eq.~(\ref{eq_5})] and slow--subdiffusion [Eq.~(\ref{eq_7})]. The latter Green function was calculated for $\eta=1.01$. The values of time and subdiffusion coefficient are given in the legend. The lines represents the Green functions for slow--subdiffusion, the symbols --- for subdiffusion. All quantities are given in arbitrarily chosen units.}
\label{Fig_1}
\end{figure}

\begin{figure}[htb]
\centerline{%
\includegraphics[scale=0.4]{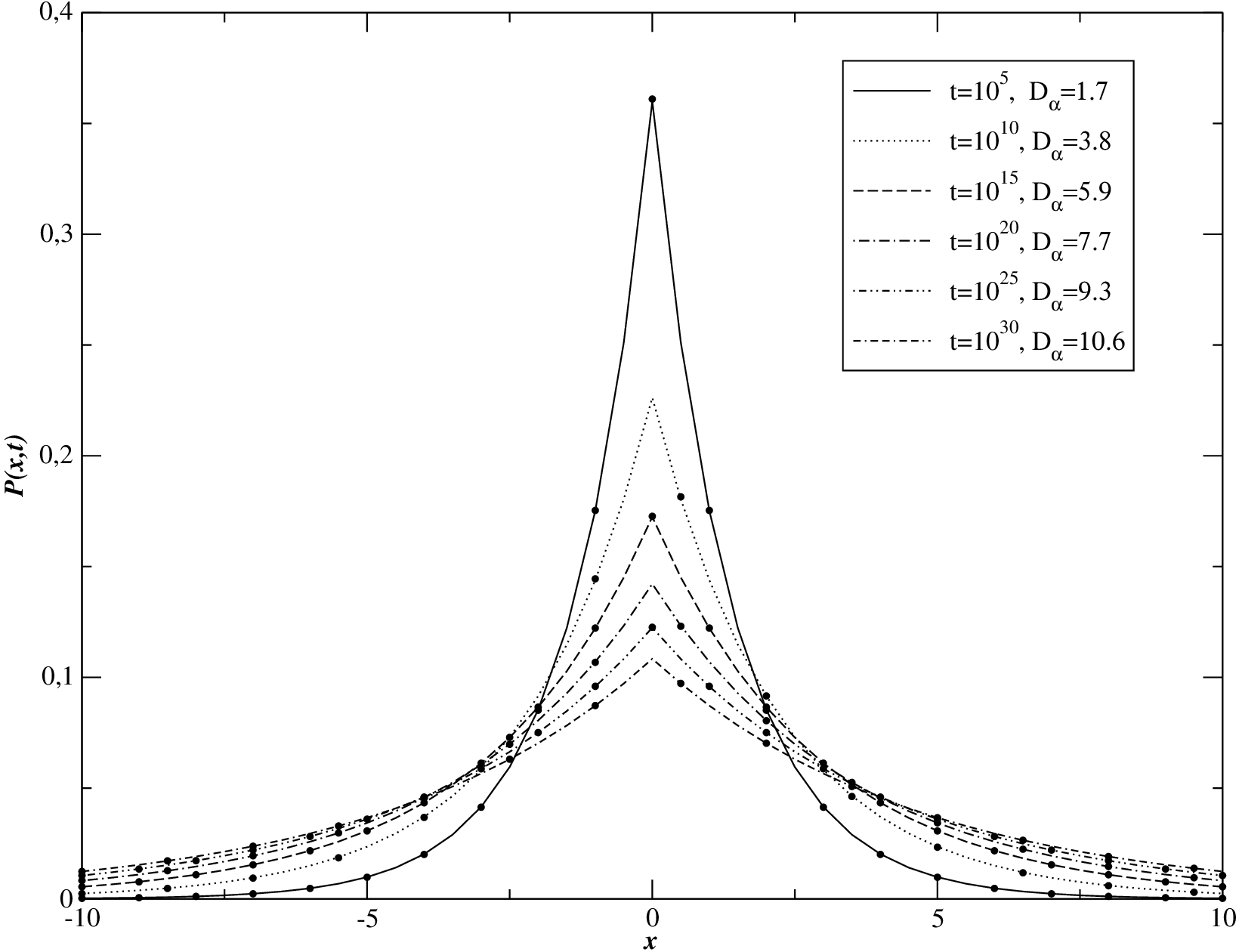}}
\caption{The same as in Fig.~\ref{Fig_1}, but for $\eta=3.54$.}
\label{Fig_2}
\end{figure}

\begin{figure}[htb]
\centerline{%
\includegraphics[scale=0.4]{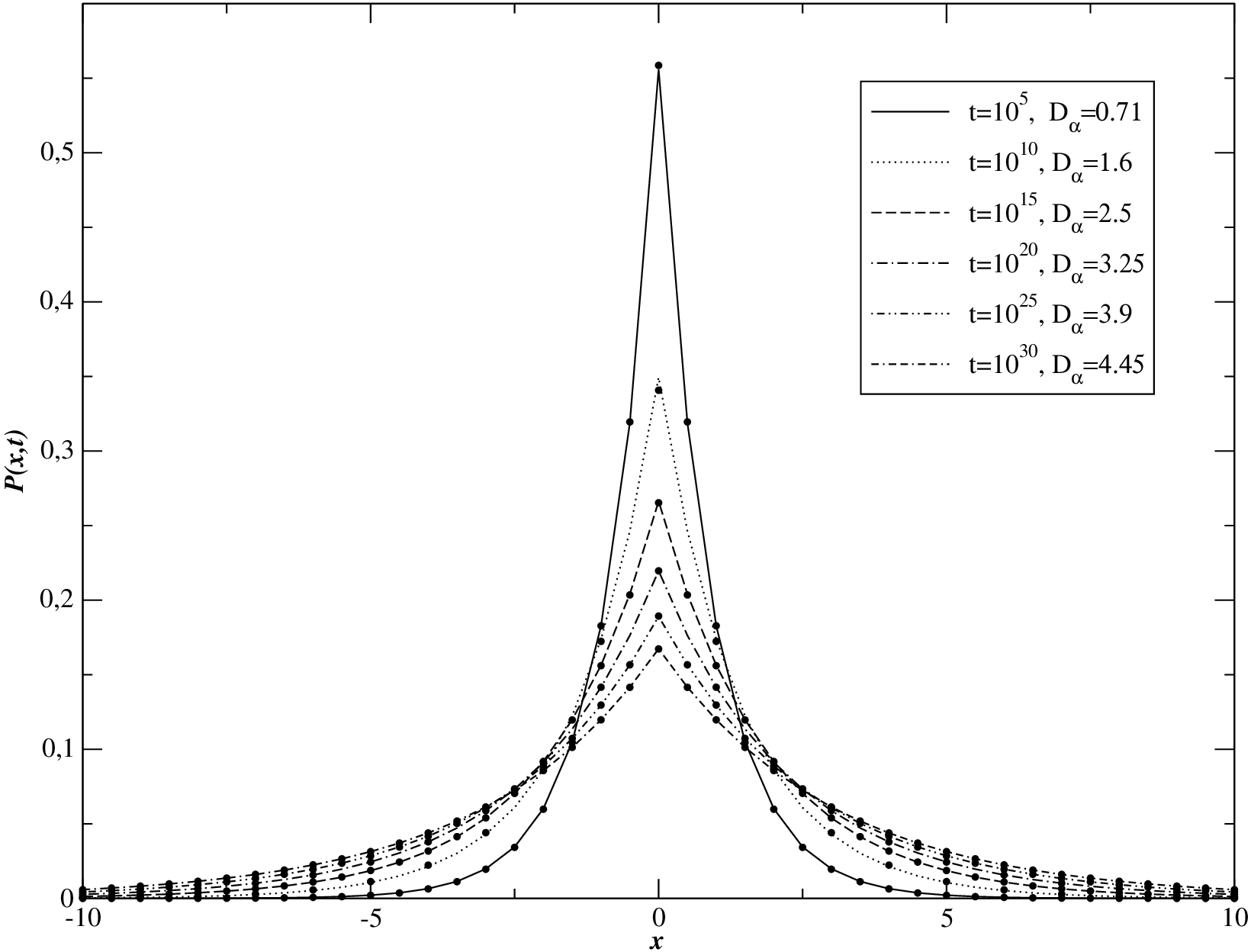}}
\caption{The same as in Fig.~\ref{Fig_1}, but for $\eta=11.23$.}
\label{Fig_3}
\end{figure}

\begin{figure}[htb]
\centerline{%
\includegraphics[scale=0.4]{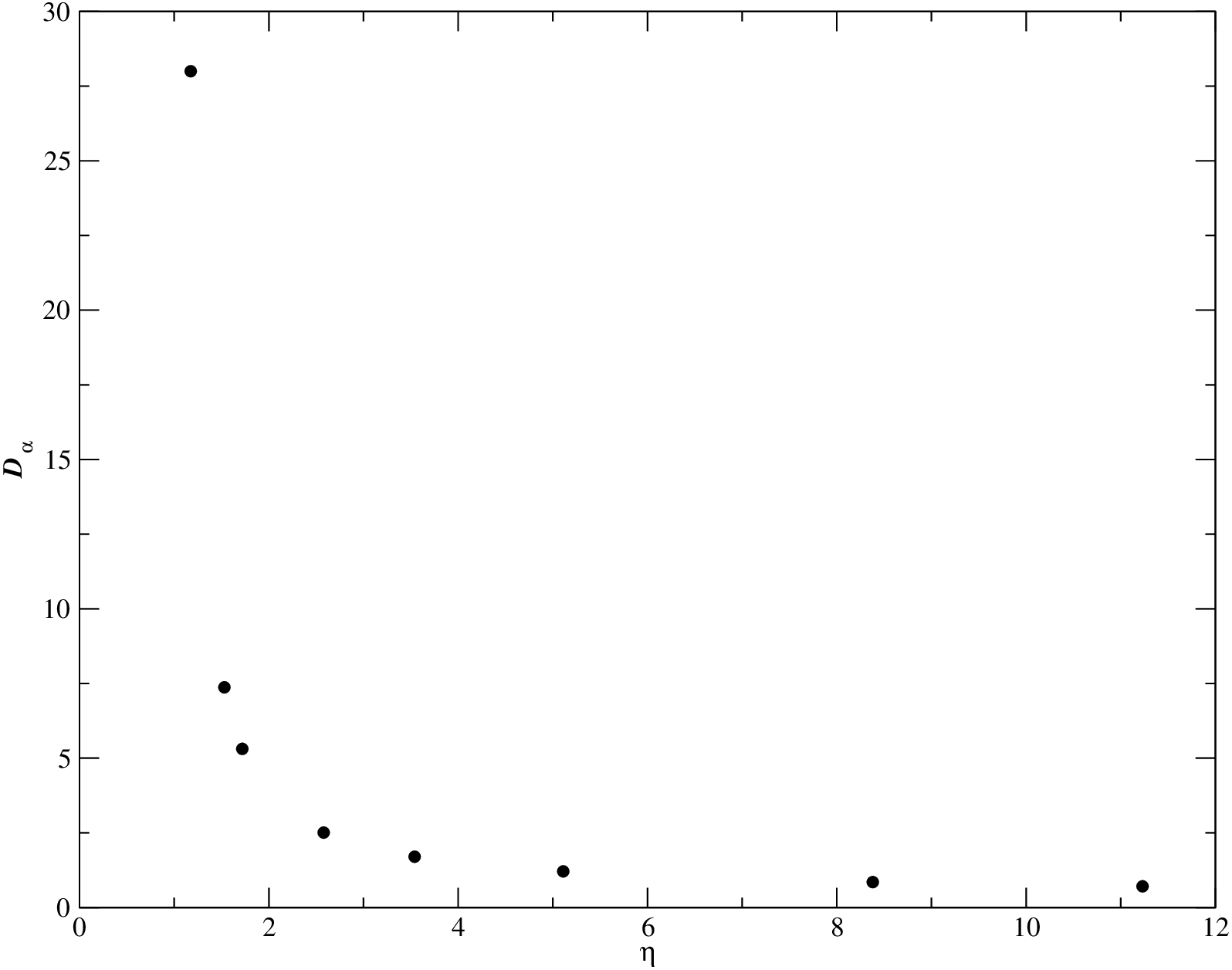}}
\caption{The dependence of the subdiffusion coefficient $D_\alpha$ on the parameter $\eta$ for $t=10^5$.}
\label{Fig_4}
\end{figure}

In Figs.~\ref{Fig_1}--\ref{Fig_3} we present the comparision between Green functions for slow--subdiffusion and subdiffusion calculates from the formulae given by Eqs. (\ref{eq_7}) and~(\ref{eq_5}), respectively. We are interested in testing if we can choose the parameters in such a way that functions (7) and (8) cover function (5). We arbitrarily chose the values of the  parameters $r$, $\sigma^2$, $\eta$ and $\alpha$. In order to match  Green functions for slow--subdiffusion and subdiffusion for all times, the subdiffusion coefficient $D_\alpha$ has to depend on time. Therefore, subdiffusion coefficient $D_\alpha$ was treated as a fitting parameter.  All calculations were performed for $r=2.34$, $\sigma^2=0.1$ and $\alpha=0.01$ and the values of the rest of the parameters were given in the legend and the figures' captions. It is interesting to find the dependance of the subdiffusion coefficient on the rest of parameters for the given value of $t$. The preliminary studies showed that the subdiffusion coefficent most strongly depends on $\eta$. In Fig.~\ref{Fig_4} we present the dependence between the fitting parameter $D_\alpha$ and the parameter $\eta$.

As can be seen in Figs.~\ref{Fig_1}--\ref{Fig_3}, Green functions for subdiffusion and  slow--subdiffusion are in quite good agreement for all values of $\eta$ for the subdiffusion coefficient $D_\alpha$ depending on time. Moreover, we can observe that the values of the subdiffusion coefficient depend strongly on the values of parameter $\eta$. In Fig.~\ref{Fig_4} we can observe that for higher values of $\eta$ we have obtained  lower values of $D_\alpha$. 
Therefore, we conclude that it is possible to describe slow--subdiffusion as subdiffusion with small values of subdiffusion parameter and a subdiffusion coefficient which depends on time. A more detailed description of the subject presented in this paper will be present elsewhere.

\section{Final reamrks}

In this paper we have compared Green functions obtained for subdiffusion and slow--subdiffusion. It seems that it is possible to descibe slow--subdiffusion by means of subdiffusion with a small value of subdiffusion parameter and the subdiffusion coefficient depending on time  although these two processes seem to be qualitatively different from each other. We have encountered a similar situation when we have approximated the solution to nonlinear sundiffusion equations by the solution to the subdiffusion equation with a fractional time derivative with the subdiffusion coefficient depending on time.

 It would be interesting to study the dependence of the subdiffusion coefficient on time or the dependence of the agreement between the Green functions for slow--subdiffusion and subdiffusion on the subdiffusion parameter $\alpha$. Moreover, investigation like this could allow one to find a more transparent physical interpretation of slow--subdiffusion.

\section*{Acknowledgments}
This paper was partially supported by the Polish National Science Centre under grant No. N N202 19 56 40 (1956/B/H03/2011/40).

\end{document}